\documentclass[prd,twocolumn,nofootinbib]{revtex4}
\usepackage{graphicx, epsfig}
\usepackage{color}
\usepackage{mathrsfs}
\usepackage{bm}
\usepackage{amsmath,amssymb}
\usepackage{mathrsfs}
\usepackage[caption=false]{subfig}
\usepackage{hyperref}
\usepackage[normalem]{ulem}
\usepackage{mathtools}
\usepackage[x11names]{xcolor}

\definecolor{fashionfuchsia}{rgb}{0.96, 0.0, 0.63}
\colorlet{no_so_fashion_purple}{blue!50!red}

\newcommand{\be}{\begin{equation}}
\newcommand{\ee}{\end{equation}}
\newcommand{\ba}{\begin{eqnarray}}
\newcommand{\ea}{\end{eqnarray}}

\newcommand{\nn}{\nonumber}

\newcommand{\hatn}{\hat n}
\newcommand{\m}{{m}}
\newcommand{\Phimm}{\Phi_{m {\bar m}}}

\newcommand{\Sinm}{\sin ( \theta_m/2 ) }
\newcommand{\Sinmbar}{\sin ( \theta_{\bar m}/2 ) }
\newcommand{\Cosm}{\cos ( \theta_m/2 ) }
\newcommand{\Cosmbar}{\cos ( \theta_{\bar m}/2 ) }

\newcommand{\zm}{d}

\begin{document}
\title{Structure of electroweak dumbbells}
\author{Teerthal Patel, Tanmay Vachaspati}
\affiliation{
$^*$Physics Department, Arizona State University, Tempe,  Arizona 85287, USA.
}

\begin{abstract}
We analyze the magnetic field of electroweak dumbbells. While the magnetic field of the
untwisted dumbbell is given by the usual dipole formula and falls of as $1/r^3$, dumbbells 
with twist have a novel twisted magnetic field that only falls off as $\cos\theta/r^2$ (in spherical
coordinates). We comment on the relevance of twisted electroweak dumbbells for understanding
the coherence of the magnetic field generated at the electroweak phase transition.
\end{abstract}

\maketitle

\section{Introduction}
\label{introduction}

The ``electroweak dumbbell'' consists of a magnetic monopole and an antimonopole of the
standard electroweak model connected by a string made of $Z-$magnetic 
field~\cite{Nambu:1977ag,Achucarro:1999it}.
The existence of such non-perturbative field configurations in the electroweak model is of 
great interest
as they would provide the first evidence for (confined) magnetic monopoles. In a cosmological 
context, dumbbells can source large-scale magnetic fields which can seed galactic magnetic 
fields and play an important role in the propagation of cosmic rays~\cite{Vachaspati:2020blt}. 

Generally electroweak dumbbells are viewed as magnetic dipoles with the usual dipolar $1/r^3$
fall off of the magnetic field strength with distance $r$ from the dipole, but the situation in
the electroweak case is richer. There is a one-parameter set of electroweak dumbbell configurations,
all describing a confined monopole-antimonopole pair but with additional structure called the 
``twist''. Such twisted dumbbells are closely related to the electroweak sphaleron as they also
carry Chern-Simons number~\cite{VachaspaticandField}. 

In the paper we investigate the structure of electroweak dumbbells using ``constrained relaxation''.
We start with a field configuration that contains a monopole and an antimonopole with 
a relative twist. The fields are then relaxed subject to the constraint that the 
orientation of the Higgs field is held fixed throughout the spatial volume, and this automatically 
fixes the monopole-antimonopole positions and the twist.

For zero twist, we find the expected dipolar structure of the magnetic field of the system. The
results for non-zero twist are more unexpected. The magnetic field lines do not connect the
monopole to the antimonopole; instead the field lines stretch to infinity, tending to pull the 
monopole and antimonopole away from each other. At large distances from the dumbbell, 
the magnetic field strength has a $\cos\theta /r^2$ behavior. The magnetic field
of a twisted dumbbell also has an azimuthal component, resembling the twisted 
field pointed out for an electroweak sphaleron in Ref.~\cite{Hindmarsh:1994ga,Hindmarsh:1993aw}.

We begin our analysis by describing the model and our initial configuration of fields 
in Sec.~\ref{model}. The numerical relaxation scheme is described in Sec.~\ref{sec:Numerical setup}.
We define the electromagnetic field in Sec.~\ref{magneticfield} and derive the magnetic field
configuration for the initial unrelaxed configuration. The results of our constrained relaxation are 
given in Sec.~\ref{results}. 
We discuss consequences of our findings in Sec.~\ref{conclusions}.

\section{Model}
\label{model}

\subsection{Electroweak model}
The Lagrangian for the bosonic sector of the electroweak theory is given by
\ba
	\mathcal{L} &&= -\frac{1}{4}W^a_{\mu\nu}W^{a\, \mu\nu} - \frac{1}{4}Y_{\mu\nu}Y^{\mu\nu} \nn \\
	&& \hskip 2.5 cm
	+ |D_\mu\Phi|^2 -\lambda(|\Phi|^2-\eta^2)^2\, ,
\ea
where
\begin{equation}
	D_\mu \equiv \partial_\mu - \frac{i}{2}g\sigma^aW^a_\mu - \frac{i}{2}g'Y_\mu\, .
\end{equation}
Here, $\Phi$ is the Higgs doublet, $W^a_\mu$ are the SU(2)-valued gauge fields with $a=1,2,3$ and, 
$Y_\mu$ is the U(1) hypercharge gauge field. In addition, $\sigma^a$ are the Pauli spin matrices
with ${\rm Tr}(\sigma^a \sigma^b) = 2 \delta_{ab}$, 
and the experimentally measured values of the parameters that we adopt from \cite{Particledatagroup2022}  are $g=0.65$, $\sin^2\theta_w = 0.22$, $g'=g\tan\theta_w$, $\lambda=0.129$ and $\eta=174 {\text GeV}$.

We adopt the temporal gauge for convenience in numerical implementation, with $W^a_0 = B_0 = 0$.
The Euler-Lagrange equations of motion for the model are given by
\ba
&&
D_\mu D^\mu \Phi + 2\lambda(|\Phi|^2  - \eta^2)\Phi = 0 \label{EL phi}\\
&&
\partial_\mu Y^{\mu\nu} = g'\, {\text Im}[\Phi^{\dagger} ( D^\nu\Phi ) ]\label{EL B}\\
&&
\partial_\mu W^{a\mu\nu} + g\epsilon^{abc} W^b_\mu W^{c\mu\nu} = 
g\, {\rm Im}[ \Phi^{\dagger} \sigma^a ( D^\nu \Phi )]
\label{EL W}
\ea
where the gauge field strengths are given by
\ba
W^{a}_{\mu\nu} &=& \partial_{\mu} W^a_\nu - \partial_{\nu} W^a_\mu + g \epsilon^{abc} W^b_\mu W^c_\nu \\
Y_{\mu\nu} &=& \partial_{\mu} Y_\nu - \partial_\nu Y_{\mu} \, .
\ea

Electroweak symmetry breaking results in three massive gauge fields, the two charged
$W$ bosons and $Z_\mu$,
and one massless gauge field, $A_\mu$, that is the electromagnetic gauge field. We define
\ba
Z_\mu &\equiv& \cos\theta_w n^a W^a_\mu + \sin \theta_w Y_\mu,  \\
A_\mu &\equiv& - \sin\theta_w n^a W^a_\mu + \cos \theta_w Y_\mu\,,
\label{A and Z field definitions unitary gauge}
\ea
where
\begin{equation}
\label{Higgs n vector}
	n^a \equiv \frac{\Phi^{\dagger}\sigma^a\Phi}{|\Phi|^2} \, .    
\end{equation}
The weak mixing angle, $\theta_w$ is given by $\tan{\theta_w} = g'/g$, the electric charge 
is given by $e = g_z \sin\theta_w \cos \theta_w$ and the $Z$ coupling is defined as 
$g_z \equiv \sqrt{g^2+g'^2}$. The Higgs, $Z$ and $W$ boson masses are given by 
$m_H \equiv 2\sqrt{\lambda}\eta =125\, {\text{GeV}}$, 
$m_Z \equiv g_z\eta/\sqrt{2} =80\, {\text{GeV}}$ 
and $m_W \equiv g\eta/\sqrt{2}=91\, {\text{GeV}}$, respectively.

\subsection{Initial field configuration}
\label{higgsconfig}

Here we setup a field configuration that describe the dumbbell. This configuration will
become our starting point for constrained numerical relaxation, a process in which the 
total energy reduces while the monopole and antimonopole are held fixed at their
initial locations.

\begin{figure}
	\centering
	\includegraphics[width=0.40\textwidth]{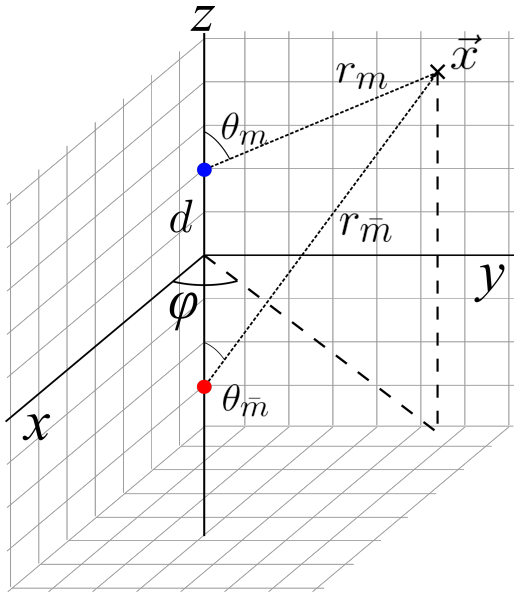}
	\caption{A general vector $ \vec{x} $ in a Cartesian 
	grid with the monopole and antimonopole centered polar angles, $\theta_m$ and $\theta_{\bar m}$, 
	respectively.  
	}
	\label{fig:illustration}
\end{figure}

\begin{figure}
\includegraphics[width=0.45\textwidth]{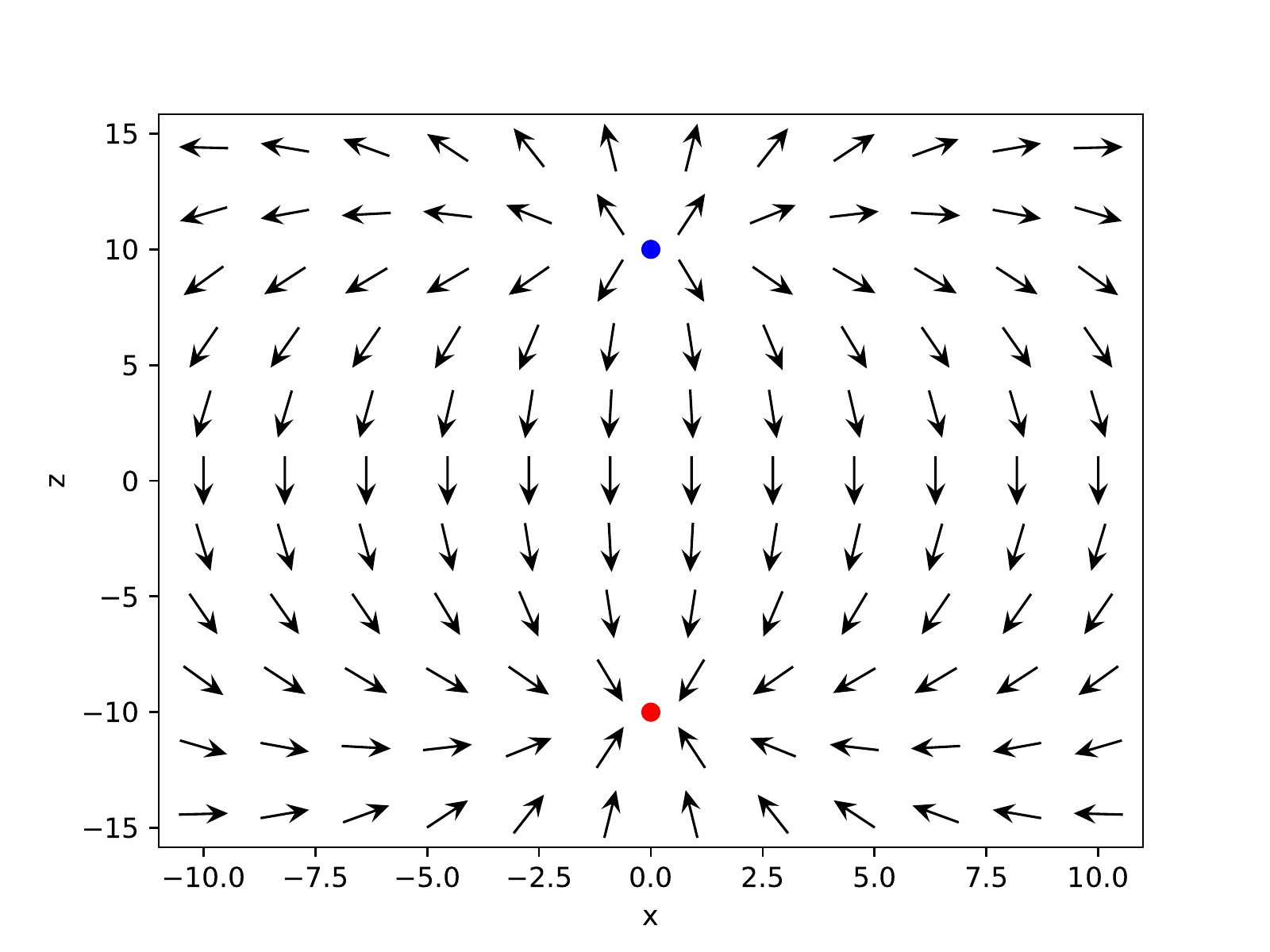}
\caption{
The projection of the vector $\hat{n}$ given by \eqref{Higgs n vector} for the 
ansatz \eqref{Phimmbar} in the $xz$ plane for $\gamma = 0$. The blue and red dots represent the 
monopole and antimonopole, respectively.}
\label{fig:.n vec twist 0}
\end{figure}

\begin{figure}
\includegraphics[width=0.45\textwidth]{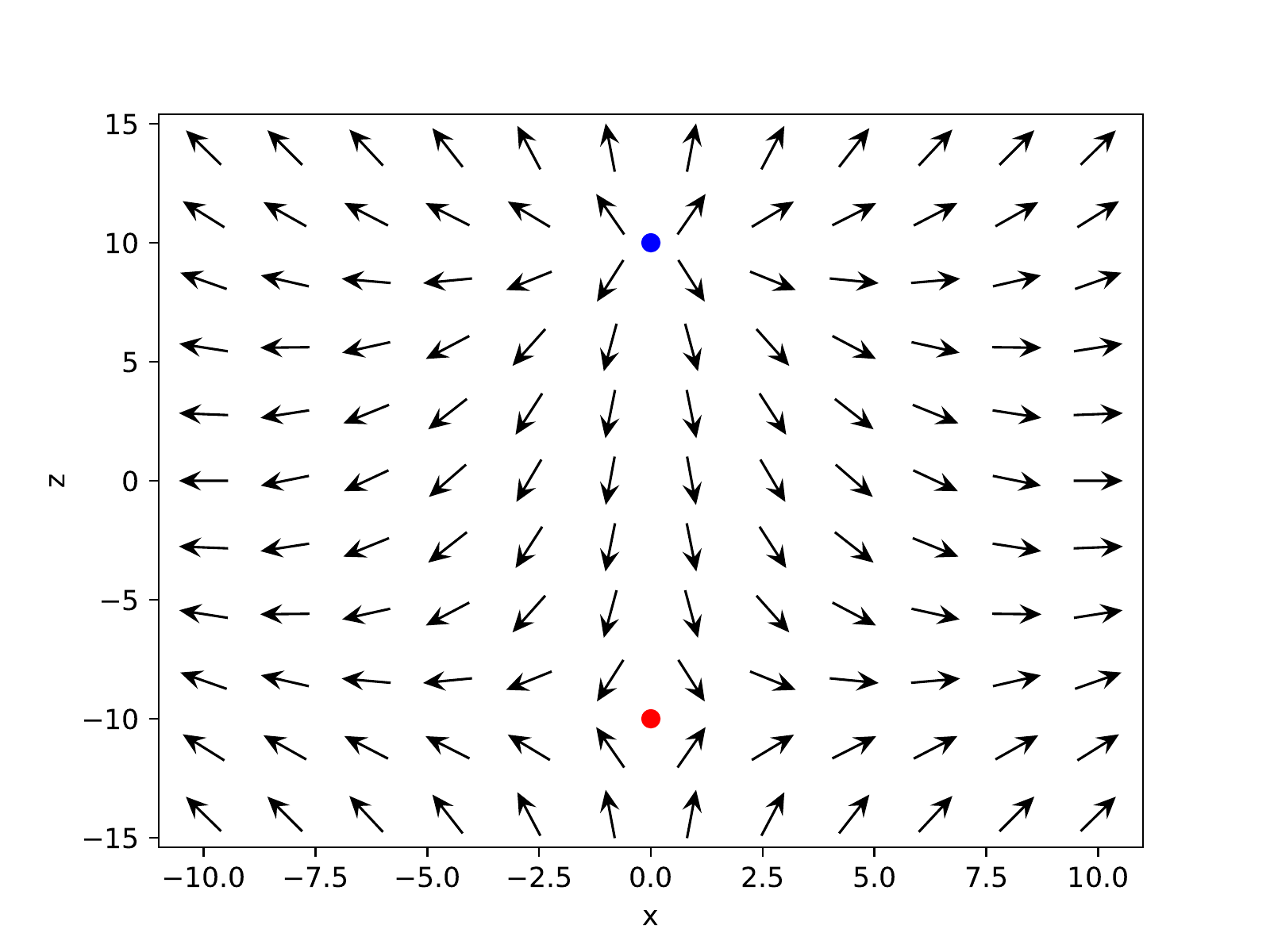}
\caption{The projection of the vector $\hat{n}$ given by \eqref{Higgs n vector} for the 
	ansatz \eqref{Phimmbar} in the $xz$ plane for $\gamma = \pi$. The blue and red dots represent 
	the monopole and antimonopole, respectively. }
\label{fig:.n vec twist pi}
\end{figure}

The angular distribution of the Higgs field for the monopole and antimonopole can be
taken to be \cite{Nambu:1977ag},
\begin{equation}
\label{Single Monopole and Antimonopole scalars}
	{\hat \Phi}_m = \left(\begin{array}{c}
		\Cosm
		\\ 
		\Sinm e^{i\phi}
	\end{array} \right), 
	{\hat \Phi}_{\bar m} = \left(\begin{array}{c}
		\Sinmbar
		\\ 
		\Cosmbar e^{i\phi}
	\end{array} \right), 
\end{equation}
where the hat on $\Phi$ denotes that $|\hat\Phi |=1$,
$(\theta_m, \phi )$ are spherical angular coordinates centered on the monopole, 
and $(\theta_{\bar{m}},\phi )$ are corresponding angles centered on the anti-monopole,
as shown in Fig.~\ref{fig:illustration}. 
The combined monopole-antimonopole ansatz for the Higgs field can be taken to 
be~\cite{VachaspaticandField},
\begin{widetext}
\begin{equation}
{{ {{\hat \Phi}_{\m{\bar m}}}}(\gamma)} = \left( 
\begin{array}{c}
\Sinm \Sinmbar e^{i\gamma} + \Cosm\Cosmbar\\
\Sinm \Cosmbar e^{i\phi} - \Cosm \Sinmbar e^{i(\phi -\gamma)}
\end{array}
\right)\, ,
\label{Phimmbar}
\end{equation}
\end{widetext}
where the monopole and antimonpole are located along the z-axis, as illustrated in Fig.~\ref{fig:illustration},
and we have included a ``twist'' angle $\gamma$. 
In the limit $ \theta_{\bar{m}}\rightarrow 0 $, \eqref{Phimmbar}, one recovers the monopole configuration 
and, in the limit $ \theta_\m \rightarrow \pi$, one recovers the antimonopole
configuration in \eqref{Phimmbar} but with a twist given by $\gamma$. 
The effect of the relative twist $ \gamma$ is more evident from the 
vector \eqref{Higgs n vector} associated with the Higgs field. This is visualized in Cartesian space 
in Figs.~\ref{fig:.n vec twist 0} and \ref{fig:.n vec twist pi}. 
Here, we see the projection $ \hat{n} - (\hat{n}\cdot\hat{j})\hat{j} $
of the vector in the $xz-$plane for extreme values of twists $ \gamma = 0,\pi$.


The gauge field configurations are obtained by setting the covariant derivative 
of the Higgs field to vanish 
in the symmetry broken regions. This procedure does not fix the gauge fields
completely as it allows for an arbitrary electromagnetic gauge field (since the
generator of the electromagnetic group annihilates $\Phi$). The electromagnetic
gauge field is defined as,
\begin{equation}
\label{EW tensor standard def}
A_{\mu} \equiv - \sin\theta_w {\hat{n}^aW^a_{\mu}} + \cos\theta_w Y_{\mu}\, .
\end{equation}
We completely fix the form of the gauge fields by requiring that $A_\mu=0$.
The gauge fields are then given by,
\ba
gW^a_\mu &=& -\epsilon^{abc}n^{b}{\partial_\mu n^c} \nn \\
&& - i \cos^2\theta_w n^a 
                           (\hat\Phi^{\dag}\partial_{\mu}\hat\Phi - \partial_{\mu}\hat\Phi^{\dagger}\hat\Phi) \\
g'Y_\mu &=& -i\sin^2\theta_w(\hat\Phi^{\dagger}\partial_{\mu}\hat\Phi - \partial_{\mu}\hat\Phi^{\dagger}\hat\Phi)
\label{MMBar gauge fields}
\ea

To correctly account for the radial dependence of the Higgs field around the monopole-antimonopole pair, we attach radial profiles. The general monopole solution is given by 
\begin{equation}
	\Phi = h(r)\hat{\Phi} \, ,
\end{equation}
where $\hat{\Phi}$ are the normalized doublets in \eqref{Single Monopole and Antimonopole scalars}.  
Including profile functions in the ansatz, the initial monpole-antimonopole scalar field configuration is 
given by
\begin{equation}
	{{\Phimm}} = k({\vec{x}})h(r_m)h(r_{{\bar{m}}}) { {{\hat \Phi}_{\m{\bar m}}}}\, ,
\label{general mmbar scalar config}	
\end{equation}
where $r_\m$ and $r_{\bar m}$ are radial coordinates centered on the monopole and antimonpole, 
respectively, given by
\begin{equation}
	r_m = |{\vec x}- {\vec x}_\m|, \,\,\,\, r_{\bar{m}} = |{\vec x} - {\vec x}_{\bar{m}} |\, ,
\end{equation}
where ${\vec x}_m = (0,0,\zm)$ and ${\vec x}_{\bar{m}} = (0,0,-\zm)$.
While $h(r)$ represents the monopole profile function,
$k({\vec{x}})$ in \eqref{general mmbar scalar config} is the profile
of the Z-string that connects the monopole and the antimonopole in the dumbbell. 

Similar to \eqref{general mmbar scalar config}, we include radial profiles for the gauge fields as
\ba
gW^a_\mu &=& l({\vec{r}})j(r_\m)j(r_{\bar{m}})[-\epsilon^{abc}n^{b}{\partial_\mu n^c} \nn \\
&& \hskip 0.5 cm
 - i \cos^2\theta_w \, n^a({\hat \Phi}^{\dagger}\partial_{\mu}{\hat \Phi} 
- \partial_{\mu}{\hat \Phi}^{\dagger}{\hat \Phi})] 
\label{Winitial} \\
g'Y_\mu &=& l({\vec{r}})j(r_\m)j(r_{\bar{m}})[ \nn \\
&& \hskip 0.5 cm
-i\sin^2\theta_w \, ({\hat \Phi}^{\dagger}\partial_{\mu}{\hat \Phi} - \partial_{\mu}{\hat \Phi}^{\dagger}{\hat \Phi})]
\label{Gauge field with radial profile}
\ea

Now that we have set up a field configuration that describes the initial dumbbell in 
Eqs.~\eqref{general mmbar scalar config}, \eqref{Winitial} and \eqref{Gauge field with radial profile},
we will perform a constrained numerical relaxation, a process in which the 
total energy reduces while the monopole and antimonopole are held fixed at their
initial positions.

\subsection{Profiles}\label{subsection:profiles}

The monopole profile functions are only known in closed form within the context of the 
Bogomolnyi-Prasad-Sommerfield (BPS) limit
($ \lambda\rightarrow0$)~\cite{PrasadandSommerfield75} . Numerical solutions for the general 
case have been outlined in \cite{Manton:2004tk}. 
 A functional form that reduces to the BPS case is given by~\cite{Vachaspati:2015ahr},
\begin{equation}
	h(r) = \frac{1}{\tanh(\eta r)} - (1+mr)\frac{e^{-mr}}{\eta r}\, ,
\label{Monopole radial profile}
\end{equation}
\begin{equation}
	j(r) = 1-\frac{\eta r}{\sinh(\eta r)} \, ,
\end{equation}
where $m = 2\sqrt{\lambda}\, \eta$ is the scalar mass and $r$ is the radial coordinate 
centered around the monopole.
In our application, we use these radial profiles as initial guess 
functions for the monopole and the antimonopole.

\begin{figure}
\includegraphics[width=0.45\textwidth]{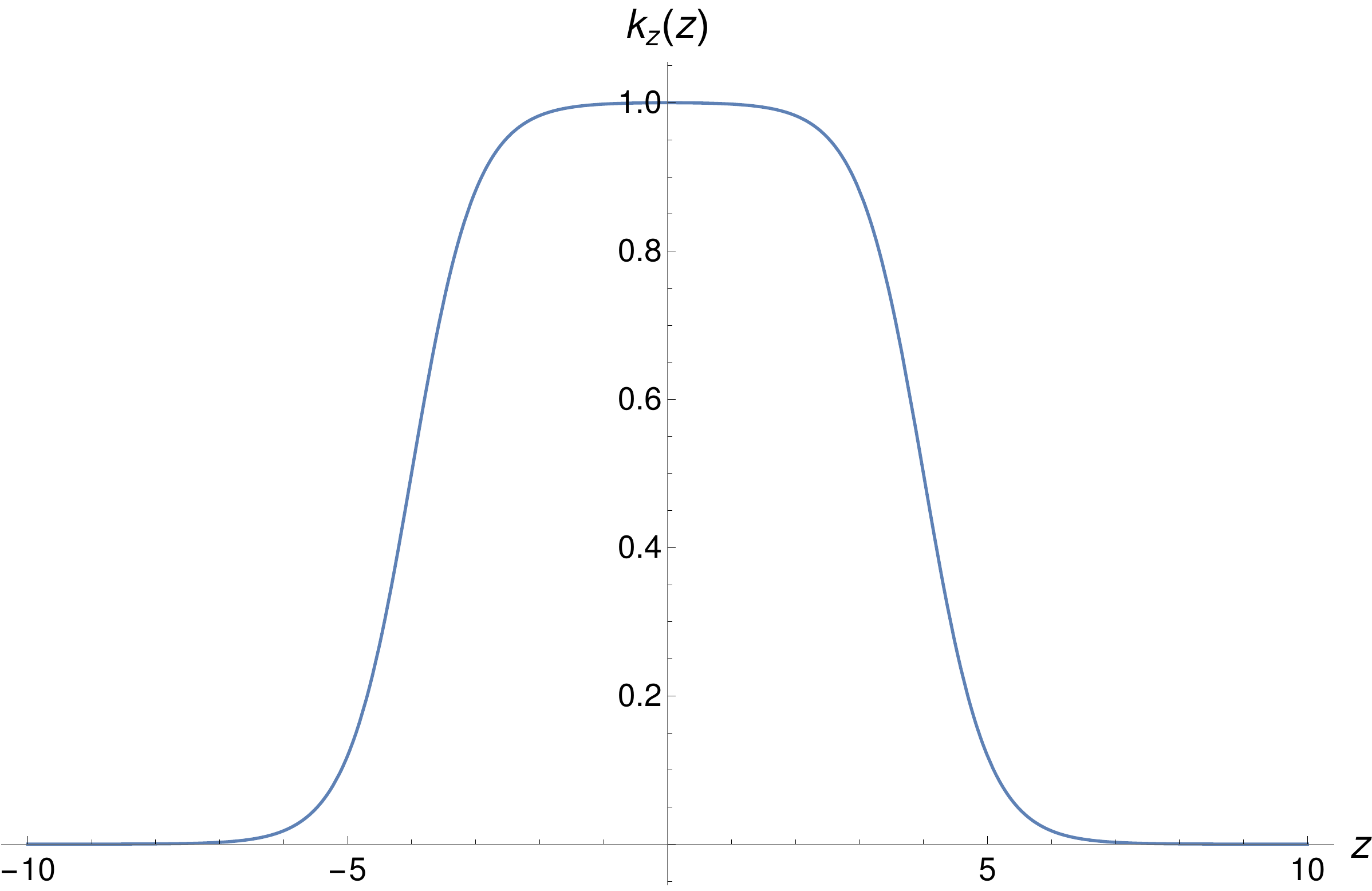}
\caption{The component $k_z(z)$  of the string profile $k(\vec{x})$, given by \eqref{k_z}, as a function of $z$. 
The monopole and antimonopole are located at $z = \pm 4\,\eta^{-1}$, respectively.
}
\label{kzplot}
\end{figure}

\begin{figure}
\includegraphics[width=0.45\textwidth]{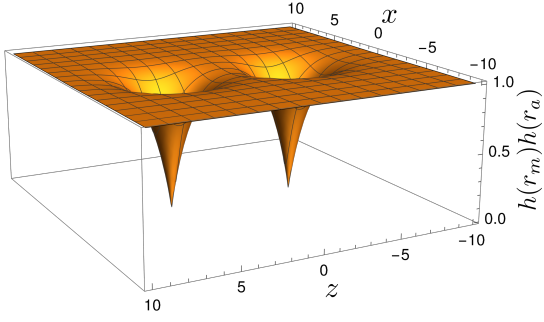}
\caption{The contour plots for the profile function $h(r_m)h(r_{{\bar{m}}})$ in the $y=0$ plane. Here, the monopole and antimonopole are located at $z= \pm 4$, respectively.
}
\label{Combined mmbar profile}
\end{figure}
\begin{figure}
\includegraphics[width=0.45\textwidth]{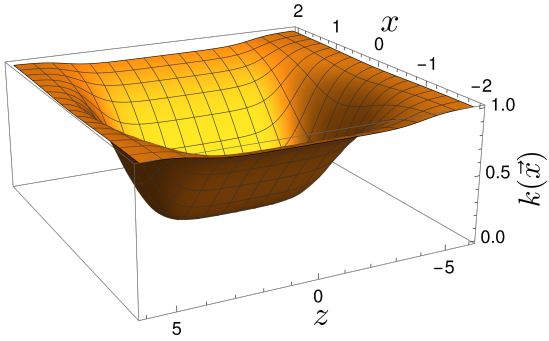}
\caption{The contour plots for the profile function $k(\vec{x})$ in the $y=0$  plane. Here, the monopole and antimonopole are located at $z= \pm 4$, respectively. 
}
\label{fig:total string profile}
\end{figure}

The string profile functions have been studied in the context of the Abelian Higgs model, known as 
Nielsen-Olesen strings~\cite{Nielsen:1973cs,Vilenkin:2000jqa}.
The Nielsen-Olesen string profiles are not known analytically, but have been studied numerically. 
Our string profile guess 
functions $ k({\vec{x}}) $ and $ l({\vec{x}}) $ match the $ \rho \rightarrow 0 $ and $ \rho \rightarrow \infty$ 
behavior of Nielsen-Olesen string solutions, where $\rho \equiv \sqrt{x^2+y^2}$. 
However, unlike the Nielsen-Olesen case, we are dealing with a finite Z-string that joins the 
monopole-antimonopole pair. 
Therefore, our guess functions depend on ${\vec{x}}$ and not just $\rho$, allowing
the string profile function to terminate smoothly at the monopole and antimonopole. 
To this end, we define the string profile function by writing it as
\begin{equation}
\label{string profile function composition}
k(\vec{x}) = 1-k_{\rho}(\rho)k_z(z)\, ,
\end{equation}
where $k_z$ is the function that ensures that the Z-string terminates at the monopole and antimonopole, 
while $k_\rho$ is the radial profile. These functions are taken to be
\begin{align}
\label{k_z}
k_z(z) &= \frac{\tanh(\eta (-z+\zm))+\tanh(\eta (z+\zm))}{2\tanh (\eta \zm)} \\
\label{k_rho}
k_\rho(\rho) &= \frac{\tanh(\eta (-\rho+\delta ))+\tanh(\eta (\rho+\delta ))}{2\tanh (\eta \delta )} \, .
\end{align}
where $\delta$ is the lattice spacing. 
$k_z$ is illustrated in Fig.~\ref{kzplot}, and the total string profile, $k(\vec{x})$, is plotted 
in Figure \ref{fig:total string profile}.

One issue is that the profile $k({\vec x})$ does not strictly vanish at the location of the string.
This is not an issue for us because in our numerical work we place the string between
lattice points and never have to evaluate the profiles exactly at the center of the string.

\section{Numerical Relaxation}
\label{sec:Numerical setup}
  
\subsection{System of equations}

The static equation of motion for $ \Phi $ can be expanded as
\ba
0&=&\partial_0^2\Phi = D_iD_i \Phi - 2\lambda(|\Phi|^2 - \eta^2)^2\Phi \nn \\
&&
= \nabla^2\Phi  - i\frac{g}{2}\sigma^aW^a_i\partial_{i}\Phi - i\frac{g'}{2}Y_i \partial_{i}\Phi \nn \\
&&
- i\frac{g}{2}\sigma^a\Gamma^a_i\Phi - i\frac{g'}{2}\Xi_i\Phi^\alpha 
-i\frac{g}{2}\sigma^aW^a_i(D_i\Phi) \nn \\
&&
- i\frac{g'}{2}Y_iD_i\Phi - 2\lambda(|\Phi|^2 - \eta^2)^2\Phi\, .
\label{Phi EOM cartesian}
\ea
Here, we introduced the notation $ \Gamma^a_i = \partial_{i}W^a_i $
and $ \Xi_i=\partial_{i}Y_i $.Since we are computing the static configurations for the monopole-antimonpole pair, 
we set all time derivatives to be 0. Similarly, the static  gauge field equations of motion lead to
\ba
\label{W Static EOM}
0&=& \nabla^2 W^a_{i} - \partial_i\Gamma^a_k
- g\epsilon^{abc} (\partial_kW^b_i)W^c_k - g\epsilon^{abc}W^b_i\Gamma^c_k\nn \\
&&  \hskip 1.5 cm
 - g\epsilon^{abc} W^b_k W^c_{ik} + g {\text{Im}}[\Phi^\dagger\sigma^a(D_i\Phi)]\,  \\ 
\label{Y Static EOM}
0 &=& \nabla^2Y_i - \partial_i\Xi_k + g'{\text{Im}}[\Phi^\dagger\sigma^a(D_i\Phi)]\, .
\ea

The relaxation scheme entails that we fix the monopole-antimonopole 
positions when solving the system of equations \eqref{Phi EOM cartesian}-\eqref{Y Static EOM}.
We implement this constraint by utilizing the gauge freedom in the model. We work
in the gauge where the Higgs directions are chosen to be $\hat\Phi_{m\bar m}$ as
given in \eqref{Phimmbar}. Since the positions of the monopole and antimonopole
are determined by the orientation of the Higgs as discussed in Ref.~\cite{Patel:2021iik}, this
fixes their positions for the entire relaxation process.
Only the magnitude of the Higgs field $|\Phi |$ and the gauge
fields, $W^a_i$ and $Y_i$ need to be relaxed to satisfy the equations of motion in 
\eqref{W Static EOM} and \eqref{Y Static EOM}. 
Thus instead of solving for the two components of the Higgs doublet, we solve for $|\Phi|$, where
\be
|\Phi | = \frac{1}{2} ( {\hat\Phi}^\dag \Phi + \Phi^\dag {\hat \Phi} ) \nn \, .
\ee
The Laplacian of $|\Phi | $ can be expressed as 
\ba\label{Laplacian mod phi expan}
	\nabla^2|\Phi| &=& \frac{1}{2}[ (\nabla^2\hat{\Phi}^\dagger)\Phi
	+ \hat{\Phi}^\dagger(\nabla^2{\Phi}) \nn \\ 
	&&  \hskip 0.5 cm
	+ (\nabla^2{\Phi}^\dagger)\hat{\Phi} +
	{\Phi}^\dagger(\nabla^2\hat{\Phi}) \nn \\
	&&  \hskip 1. cm
	+ 4 |\Phi | \partial_i \hat\Phi^\dag \partial_i \hat\Phi
\ea
where we have used
\be
\partial_i \Phi = \partial_i |\Phi |\, \hat\Phi + |\Phi | \partial_i\hat\Phi .
\ee
and $\partial_i (\hat\Phi^\dag \hat\Phi ) =0$.
An equation of motion for $|\Phi|$ can thus be derived by using $\nabla^2{\Phi}$ 
and $\nabla^2{\Phi}^\dagger$ from \eqref{Phi EOM cartesian} 
and, using \eqref{Phimmbar} to obtain $\hat\Phi$ and its derivatives
since, as explained above, $\hat\Phi$ is held fixed throughout the relaxation.
Since, we are working in the temporal gauge, the equations for $\Gamma^a_i$ and $\Xi_i$ are trivial.

The initial Higgs and gauge 
fields are given by the configurations in Sec.~\ref{higgsconfig} and the guess profiles in 
Sec.~\ref{subsection:profiles}.
Furthermore, we evaluated the exact analytic first and second order derivatives of the Higgs and gauge fields
for the initial configuration.

\subsection{Relaxation algorithm}

We relax the initial field configuration in a cubic lattice.
At any given lattice point, the discretized system of equations is given by 
$ \mathbf{E}_\alpha[{f}] = 0 $, where $ \mathbf{E}_\alpha[{f}] $ denote
the discretized static equations for $f \in \{|\Phi |, W^a_i, Y_i\}$. 
The equations are of the wave equation type
\be
 \mathbf{E}_\alpha[{f}] = -\nabla^2 f_\alpha + S_\alpha [f] =0
\ee
where $S_\alpha$ denotes various other terms.

To illustrate the principle of numerical relaxation, consider the 
discretized second order spatial derivative,
\ba
\nabla^2 f_\alpha &\rightarrow& -6\frac{f_\alpha}{\delta^2} 
 + \frac{1}{\delta^2} [ f_\alpha(i+1,j,k) + f_\alpha(i-1,j,k) \nn \\
&&\hskip 1.5 cm 
+ f_\alpha(i,j+1,k) + f_\alpha(i,j-1,k) \nn \\
&&\hskip 1.5 cm
+ f_\alpha(i,j,k+1) + f_\alpha(i,j,k-1)] \nn \\
&& \equiv -6\frac{f_\alpha}{\delta^2} + \frac{\Delta_\alpha}{\delta^2}
\ea
and so the equation of motion may be written as
\ba
f_\alpha (i,j,k) = \frac{\Delta_\alpha}{6} -\frac{\delta^2}{6} S_\alpha
\ea
which can also be written as,
\begin{equation}\label{discretized lap expan}
	f_\alpha(i,j,k) = \frac{\delta^2}{6}\mathbf{E}_\alpha[{f_\beta}]+f_\alpha(i,j,k)\, .
\end{equation}  
The relaxation scheme is to take the left-hand side at the $n^{\rm th}$ iteration step
and the right-hand side at the previous iteration step,
\begin{equation}
	f^{(n)}_\alpha(i,j,k) = \frac{\delta^2}{6}\mathbf{E}_\alpha[{f^{(n-1)}}]+f^{(n-1)}_\alpha(i,j,k)	\, .
\end{equation}

Numerically, we iterate 
over the lattice points in a linear order to update the field values. In our setup, the updated field values 
are made immediately available for computing the field values at the next lattice site in the computation. 
This is the Gauss-Seidel method and thus does not require an additional array to hold field values
from the previous iteration.
We continue iterating while the total energy of the configuration keeps decreasing.
In the example stated above, we used first order central finite differences which results 
in the $1/6$ coefficient. 
In our numerical runs, we use sixth order finite central differences and then the coefficient 
is $6/49$.

\subsection{Numerical setup}

For most of our numerical runs, we use a $740^3$ lattice with lattice spacing $\delta = 0.05\, \eta^{-1}$.
We work in units of $\eta$.
Thus a unit of energy in our simulation corresponds to $\eta=174$~GeV.
The monopole and string radii are comparable to the inverse of the gauge boson
masses, $m_W^{-1} = \sqrt{2} \, \eta^{-1}/g \approx 44\, \delta$. Therefore there are about 44 lattice
points that resolve the radius of the monopole and a similar number for the string.

The monopole and antimonopole are zeros of the Higgs field at their respective centers. 
In addition, the coordinates 
along which the Z-string is centered are also zeros of the Higgs field. These could lead to 
artificial numerical 
singularities and so, we offset the center of the monopole and antimonopole away from the 
$z$-axis in the $xy$ plane; 
that is to say that the monopole and antimonopole are at the coordinates 
$x=y=\delta/2, z=\pm(d+\delta/2)$. 
This also implies that the Z-string lies parallel to the $z$-axis, along
$x = y = \delta/2$.
As the algorithm iterates over the lattice, the system undergoes relaxation and slowly approaches 
the desired asymptotic solution with the changes becoming infinitesimal, as the number of iterations 
performed increases.
Since this can prove to be impractical, we introduced an additional constraint that the iterations 
are stopped once the consecutive fractional difference of the total energy drops below a certain threshold. 
This threshold was taken to be $10^{-6}$.
Note that the number of total iterations required to reach a satisfactory asymptotic scaling depends 
on both simulation parameters, like the lattice size and spacing,
as well as model parameters, such as separation and twist.
We run our algorithm for a range of separations and twists. For the set of tested parameters, the 
number of iterations performed ranges between $\sim 20000 - 70000$.

\section{Magnetic field}
\label{magneticfield}

We adopt the definition for the electromagnetic field strength tensor in the
symmetry broken phase ($|\Phi|=\eta$) \cite{tHooft:1974kcl,Vachaspati:1991nm},
\ba
\label{A field def proper}
A_{\mu\nu} &\equiv& - \sin\theta_w \hatn^a W^a_{\mu\nu} + \cos \theta_w Y_{\mu\nu }\nn \\
&&
-i\frac{2\sin\theta_w}{g\eta^2}(D_\mu \Phi^\dagger D_\nu \Phi - D_\nu \Phi^\dagger D_\mu \Phi)\nn \\
&=&\partial_\mu A_\nu-\partial_\nu A_\mu\nn \\
&&
-i\frac{2\sin\theta_w}{g\eta^2}(\partial_\mu \Phi^\dagger \partial_\nu \Phi - \partial_\nu \Phi^\dagger \partial_\mu \Phi)\,.
\label{Amunu}
\ea
This definition implies the presence of non-zero electromagnetic fields for $A_\mu=0$ 
due to the Higgs gradient term. In the unitary gauge, the Higgs gradient term vanishes 
and one recovers the standard expression for the Maxwellian electromagnetic fields.

We now obtain analytic expressions for the magnetic field of the unrelaxed electroweak dumbbell
for arbitrary twist. The relaxation procedure will change the detailed features of
the magnetic field but still preserves the qualitative features of the unrelaxed configuration
as we will see in Sec.~\ref{results}.

Eqs.~\eqref{Winitial} and \eqref{Gauge field with radial profile}
ensure that $A_\mu =0$ and it is only necessary to evaluate the last term in \eqref{Amunu}.
The definition of $A_{\mu\nu}$ assumes that electroweak symmetry is broken and the
expression in \eqref{Amunu} applies in regions where $|\Phi| \approx \eta$. Hence we
can replace $\Phi$ by $\Phi_{m\bar m}$ of \eqref{Phimmbar} and write,
\be
\partial_i \Phi \equiv A \partial_i \theta_m + B \partial_i \theta_{\bar m} + C\partial_i \phi
\ee
where $A$, $B$ and $C$ are derivatives of $\Phi$ with respect to $\theta_m$,
$\theta_{\bar m}$ and $\phi$ respectively.
Then
\ba
\partial_{[i} \Phi^\dag \partial_{j]} \Phi &=&
(A^\dag B - B^\dag A) \, \partial_{[i} \theta_m  \partial_{j]} \theta_{\bar m}
\nn \\ && \hskip -0.25 cm
+(A^\dag C - C^\dag A) \, \partial_{[i} \theta_m  \partial_{j]} \phi
\nn \\ && \hskip -0.25 cm
+(B^\dag C - C^\dag B) \, \partial_{[i} \theta_{\bar m}  \partial_{j]} \phi
\ea
and the square brackets in the indices denote antisymmetrization.

From Fig.~\ref{fig:illustration} we see,
\be
\tan\theta_m = \frac{\rho}{z-\zm}, \ \ 
\tan\theta_{\bar m} = \frac{\rho}{z+\zm}.
\ee
where $\rho$ is the cylindrical radial coordinate.
Therefore, after some algebra,
\be
\partial_{[i} \theta_m  \partial_{j]} \theta_{\bar m} = 
\frac{2\zm \rho}{r_m^2 r_{\bar m}^2} \, \partial_{[i} \rho \partial_{j]} z
\ee
\be
\partial_{[i} \theta_m  \partial_{j]} \phi =
\frac{1}{r_m^2} \left [ (z-\zm) \, \partial_{[i}\rho \partial_{j]} \phi - \rho \, \partial_{[i}z \partial_{j]} \phi \right ]
\ee
\be
\partial_{[i} \theta_{\bar m}  \partial_{j]} \phi =
\frac{1}{r_{\bar m}^2} \left [ (z+\zm) \, \partial_{[i}\rho \partial_{j]} \phi - \rho \, \partial_{[i}z \partial_{j]} \phi \right ]
\ee

Then from \eqref{Amunu} we get the (cylindrical) components of the magnetic field,
\ba
B_\rho &=& -i \kappa  \left [ 
\frac{A^\dag C-C^\dag A}{r_m^2} + \frac{B^\dag C-C^\dag B}{r_{\bar m}^2} \right ]
\label{Brho}
\\ 
B_\phi &=&  i\kappa \frac{2\zm \rho}{r_m^2 r_{\bar m}^2} (A^\dag B-B^\dag A)
\label{Bphi}
\\
B_z &=&  -i \frac{\kappa}{\rho} \biggl [ (z-\zm) \frac{A^\dag C-C^\dag A}{r_m^2} \nn \\
&& \hskip 2 cm
+ (z+\zm)  \frac{B^\dag C-C^\dag B}{r_{\bar m}^2} \biggr ]
\label{Bz}
\ea
with $\kappa \equiv 2\sin\theta_w / g$.

The factors with the doublets $A$, $B$ and $C$ can be evaluated using derivatives
of \eqref{Phimmbar} with respect to $\theta_m$, $\theta_{\bar m}$ and $\phi$. The
calculation simplifies if we write
\be
\Phi = \sin(\theta_m/2) e^{i\gamma} \Phi_1 + \cos(\theta_m/2) \Phi_2
\ee
where
\be
\Phi_1 = \begin{pmatrix}
\sin (\theta_{\bar m}/2) \\ \cos (\theta_{\bar m}/2 ) e^{i(\phi-\gamma)}
\end{pmatrix},
\ee
\be
\Phi_2 = \begin{pmatrix}
\cos (\theta_{\bar m}/2) \\ -\sin (\theta_{\bar m}/2 ) e^{i(\phi-\gamma)}
\end{pmatrix}
\ee
with the properties $|\Phi_1|=1=|\Phi_2|$, $\Phi_1^\dag \Phi_2=0$,
$\Phi_1 =-2 \partial_{\theta_{\bar m}} \Phi_2$, and
$\Phi_2 = 2 \partial_{\theta_{\bar m}} \Phi_1$.
This gives
\be
A^\dag B-B^\dag A = \frac{i}{2} \sin\gamma \, \cos\theta_m . \nn
\ee
Similar calculations give
\be
A^\dag C-C^\dag A = 
\frac{i}{2} \left [ \sin\theta_m \cos\theta_{\bar m}
- \cos\gamma \cos\theta_m \sin\theta_{\bar m} \right ] ,\nn
\ee
\be
B^\dag C-C^\dag B =
\frac{i}{2} \left [ \cos\theta_m \sin\theta_{\bar m}
- \cos\gamma \sin\theta_m \cos\theta_{\bar m} \right ] ,\nn
\ee
These expressions can now be inserted in \eqref{Brho}, \eqref{Bphi}
and \eqref{Bz}. The resulting expressions are not transparent and
we shall focus on a few interesting features.

First of all consider the magnetic field twist given by $B_\phi$,
\be
B_\phi =  -\kappa \zm  \sin\gamma \frac{\rho}{r_m^2 r_{\bar m}^2} \cos\theta_m
\ee
Note that the twisting is not symmetric in $\theta_m$ and $\theta_{\bar m}$.
For example, $B_\phi$ vanishes for $\theta_m=\pi/2$ but not when
$\theta_{\bar m}=\pi/2$.
This is a feature stemming from our choice of the unrelaxed field. Upon relaxation
the twist gets redistributed as in apparent in Sec.~\ref{results} where
we plot the magnetic field in the $\theta_m=\pi/2$ plane. Also note that
the twisting reverses direction under $\gamma \to -\gamma$.

Far from the dumbbell $r_m \approx r_{\bar m} \approx r$,
$\theta_m \approx \theta_{\bar m} \approx \theta$, and $\rho =r \sin\theta$.
Then
\be
B_\phi =  -\kappa \zm \sin\gamma \frac{\sin\theta \cos\theta }{r^3} 
\label{Bphiasymp}
\ee
Hence the azimuthal field is non-vanishing only for $\gamma\ne 0$, falls off as $1/r^3$,
and the twisting is of opposite signs for $\cos\theta >0 $ and $\cos\theta <0$.

Next we consider the magnetic field in the $z=0$ plane. Then we have
$r_m = r_{\bar m}=\rho$ and $\theta_m+\theta_{\bar m} = \pi$. This gives us
\be
B_\rho (z=0) = \frac{\kappa}{2\rho^2} (1 - \cos\gamma )\sin(\theta_m+\theta_{\bar m}) = 0
\ee
\be
B_z (z=0) =  \frac{\kappa \zm}{2\rho^3} \sin(2\theta_m) \, (1+ \cos\gamma ).
\ee
Hence for maximal twist angle, $\gamma = \pi$, the magnetic field on the $z=0$
plane vanishes.

Finally we consider the asymptotic magnetic field. We have already calculated the
azimuthal component in \eqref{Bphiasymp}. For the other components, note that
once again $r_m \approx r_{\bar m} \approx r$, $\theta_m \approx \theta_{\bar m} \approx \theta$ 
and so
\be
B_\rho \bigl |_{r \gg \zm} = \kappa (1 - \cos\gamma )\frac{\sin\theta \cos\theta}{r^2}
\label{Brholarger}
\ee
and
\be
B_z \bigl |_{r \gg \zm}  =   \kappa (1-\cos\gamma)   \frac{\cos^2\theta}{r^2}
\label{Bzlarger}
\ee
Note that the magnetic field of the twisted dumbbells ($\gamma \ne 0$) falls off as $1/r^2$,
instead of the dipolar $1/r^3$. 

Some more insight is gained by calculating the spherical radial component of the 
magnetic field in the asymptotic region. The radial component is given by
\be
B_r = B_\rho \sin\theta + B_z \cos\theta.
\ee
Using \eqref{Brholarger} and \eqref{Bzlarger} we get
\be
B_r \bigl |_{r \gg \zm} = \kappa (1 - \cos\gamma )\frac{\cos\theta}{r^2}.
\label{Brlarger}
\ee
Therefore the radial field has the structure of a {\it monopole's} 
magnetic field that has been squeezed into the angular range $0 \le \theta \le \pi/2$, and 
an {\it antimonopole's} magnetic field squeezed in the angular range $\pi/2 < \theta < \pi$.
The magnetic field vanishes at $\theta = \pi/2$. The long range magnetic field of a 
twisted dumbbell has a $1/r^2$ fall off, like that of a monopole. Only in the untwisted
($\gamma=0$) case does this monopole contribution vanish, and then the
dipole $1/r^3$ term becomes the leading contribution.

Many of the qualitative features of the initial magnetic field persist even after relaxation as we
now discuss.

\section{Results}
\label{results}

As a check of our numerical relaxation scheme we have calculated energies of the electroweak Z-string
\cite{PhysRevLett.68.1977} and the electroweak sphaleron.
Our relaxation procedure on a three dimensional lattice gives the energy per unit length of the Z-string 
to be $1.023\pi\eta^2$. This is within $2\%$ of the values previously calculated from numerical solutions 
of the radial differential equations~\cite{Nielsen:1973cs,Vilenkin:2000jqa}. 
For the second check, we obtained the energy of the electroweak sphaleron by using the 
configuration in \eqref{Phimmbar} with twist $\gamma=\pi$ and zero separation, {\it i.e.}
$\theta_m = \theta_{\bar m}$. Then $\Phi_{m\bar m}$ has the configuration of the 
$SU(2)$ ($\theta_w=0$) sphaleron. On relaxing the configuration, we find
the sphaleron energy to be $2.00 \times 4\pi$ for $\lambda=1/2$ and $\theta_w=0$, 
which is within $1\%$ of the result in \cite{Manton:2004tk}. 
\begin{figure}
	\centering
	\includegraphics[width=0.5\textwidth]{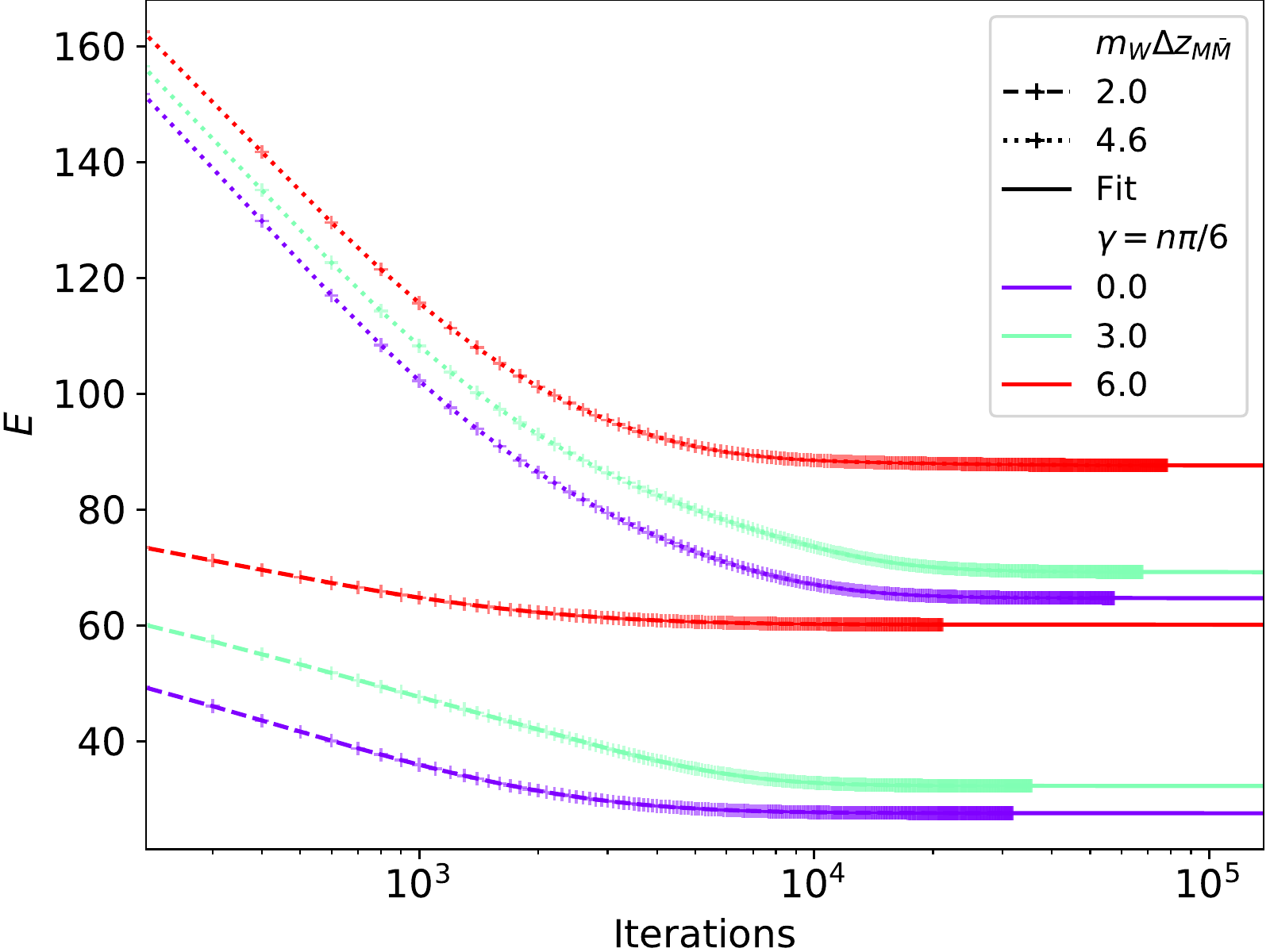}
	\caption{The energy of the configuration as a function of the number of iterations for twist angle 
	$n\pi/6$, $n=0,3,6$. The different colors and line styles correspond to different values of $n$ and 
	monopole-antimonopole separation $\Delta z_{M\bar M}$, as indicated in the legends. The solid 
	lines are asymptotic fits to the converged energies.}
	\label{fig:energyoveriterations}
\end{figure}

\begin{figure}
	\centering
	\includegraphics[width=0.5\textwidth]{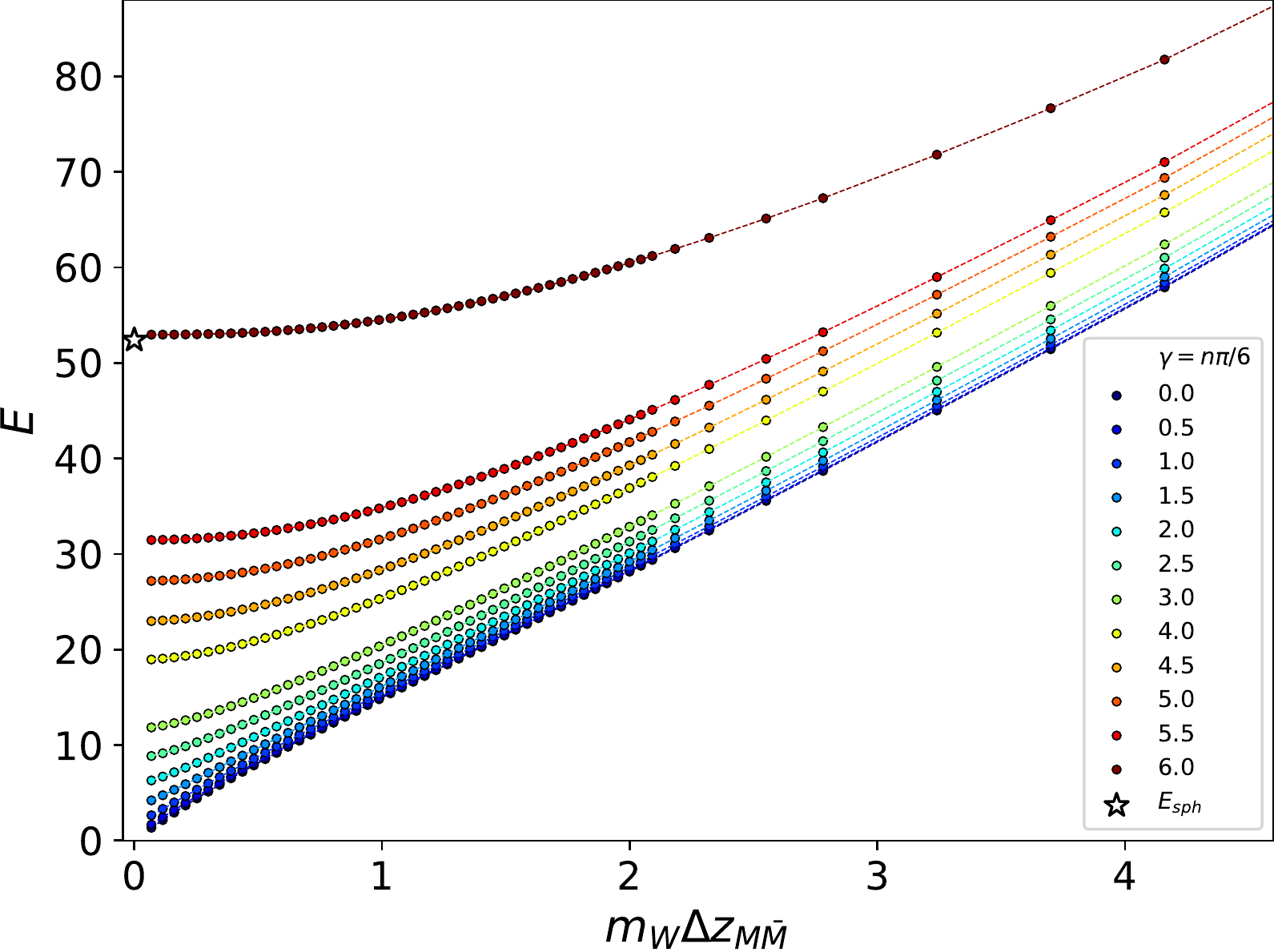}
	\caption{The energy of electroweak dumbbells as a function of monopole-antimonopole
	separation for twist angle $\gamma = n\pi/6$ for various values of $n$ as shown in the
	legend.
	The energies increase as we increase
	the twist. The star at minimal separation and $\gamma=\pi$ denotes the sphaleron.
}
	\label{fig:energysep}
\end{figure}

In Fig.~\ref{fig:energyoveriterations}, we show the energies of electroweak dumbbell
configurations for two different monopole-antimonopole separations, and a few different 
twists as a function of the number of iterations in our relaxation procedure. The solid lines 
are asymptotic fits that show convergence of the energies of the configurations. In Fig.~\ref{fig:energysep} we show the energies of the relaxed electroweak 
dumbbells for twist angle $\gamma \in [0,\pi]$, and for a range of separations. 
At large separations, most of the energy is in the string and hence
we see linear growth. At small separations the string is less important and monopole-antimonopole 
interactions become important. The flattening of the $\gamma=\pi$ curve at very small separations
indicates the existence of an unstable solution, which is precisely the sphaleron.

\begin{figure}
	\centering
	\includegraphics[width=0.5\textwidth]{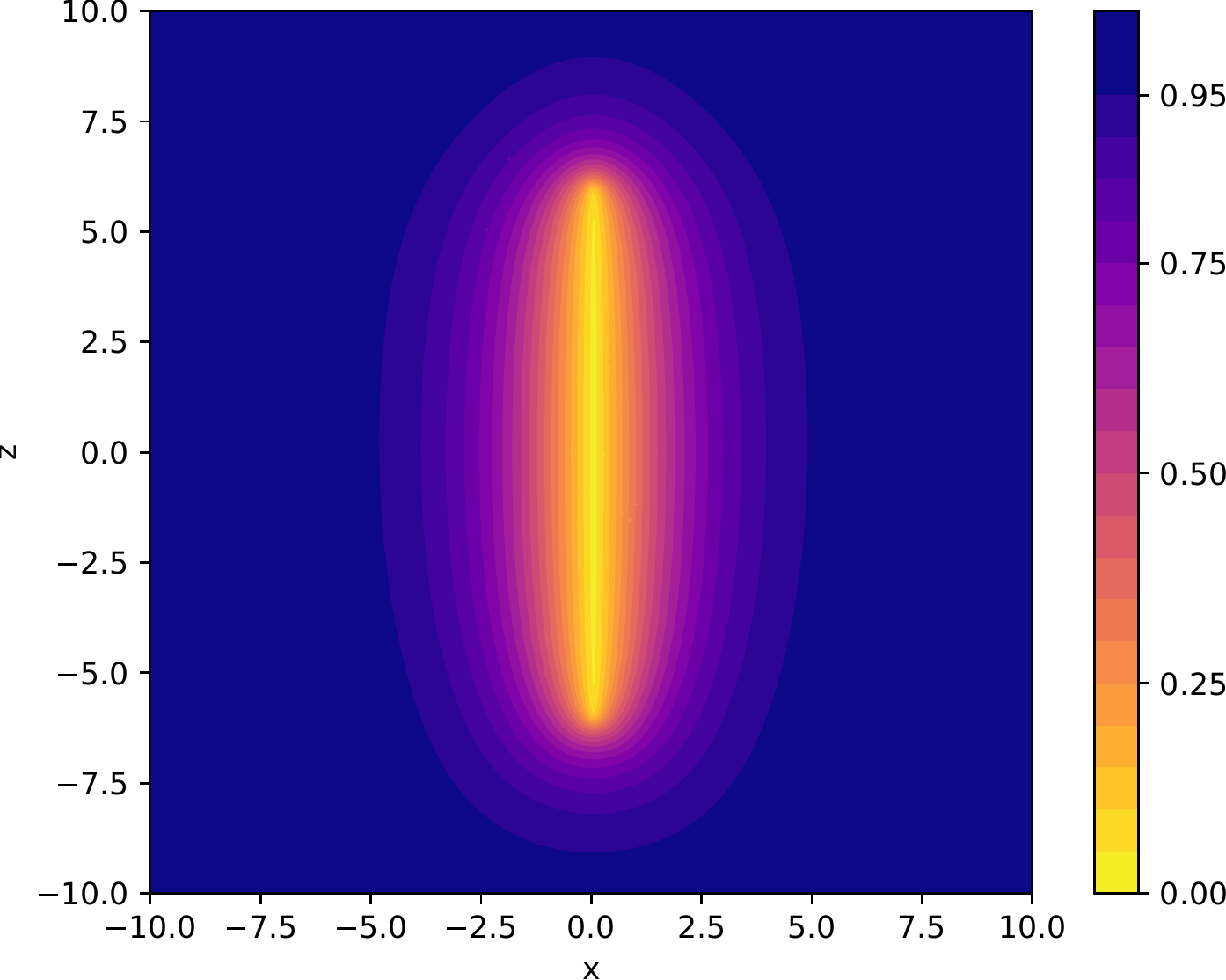}
	\caption{Contours of the Higgs field magnitude $|\Phi|$ in units of $\eta$ in the $xz$-plane after relaxation for $\gamma=\pi$ and monopole-antimonopole separation of $2d=120\delta=12\eta^{-1}$. The same contour for the $\gamma=0$ case has very similar features.}
	\label{fig:xzphicontours60000twistpi}
\end{figure}

In Fig.~\ref{fig:xzphicontours60000twistpi}, we
show contours of the magnitude of the relaxed Higgs field $|\Phi|$ in the $xz$-plane for twist 
$\pi$. There are no significant differences in features for different values of the twist $\gamma$.


\begin{figure}
\includegraphics[width=0.5\textwidth]{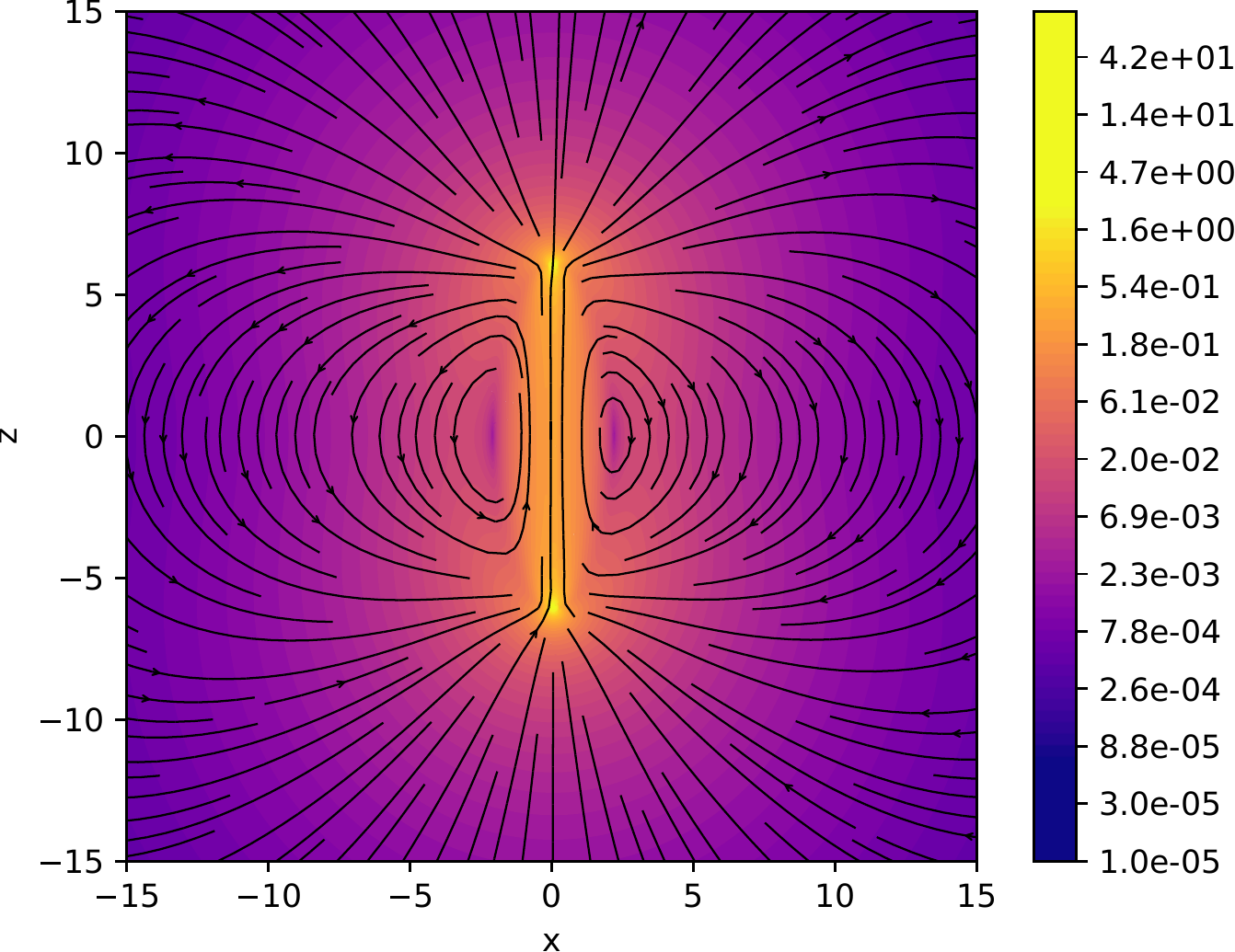}
\caption{
Contours of the magnetic field magnitude $ |B| $ (colors) in units of $\eta^2$, and the projection 
of magnetic field lines in the $xz-$plane after relaxation for $\gamma=0$, and
monopole-antimonopole separation $2\zm = 120\delta = 12 \eta^{-1}$.
}
\label{fig:xz twist 0 mag fields}
\end{figure}

To find the magnetic structure of the electroweak dumbbell we use the definition of the
field strength in \eqref{Amunu}. This expression assumes $|\Phi|=\eta$ and hence is strictly
valid only far from the dumbbell. However, we will apply it to the entire volume; points where
$\Phi=0$ are avoided since the dumbbell zeros are situated between lattice points.
In Fig.~\ref{fig:xz twist 0 mag fields} we show the magnetic field strength (colors) and
the magnetic field lines of an untwisted dumbbell. 

\begin{figure}
\includegraphics[width=0.5\textwidth]{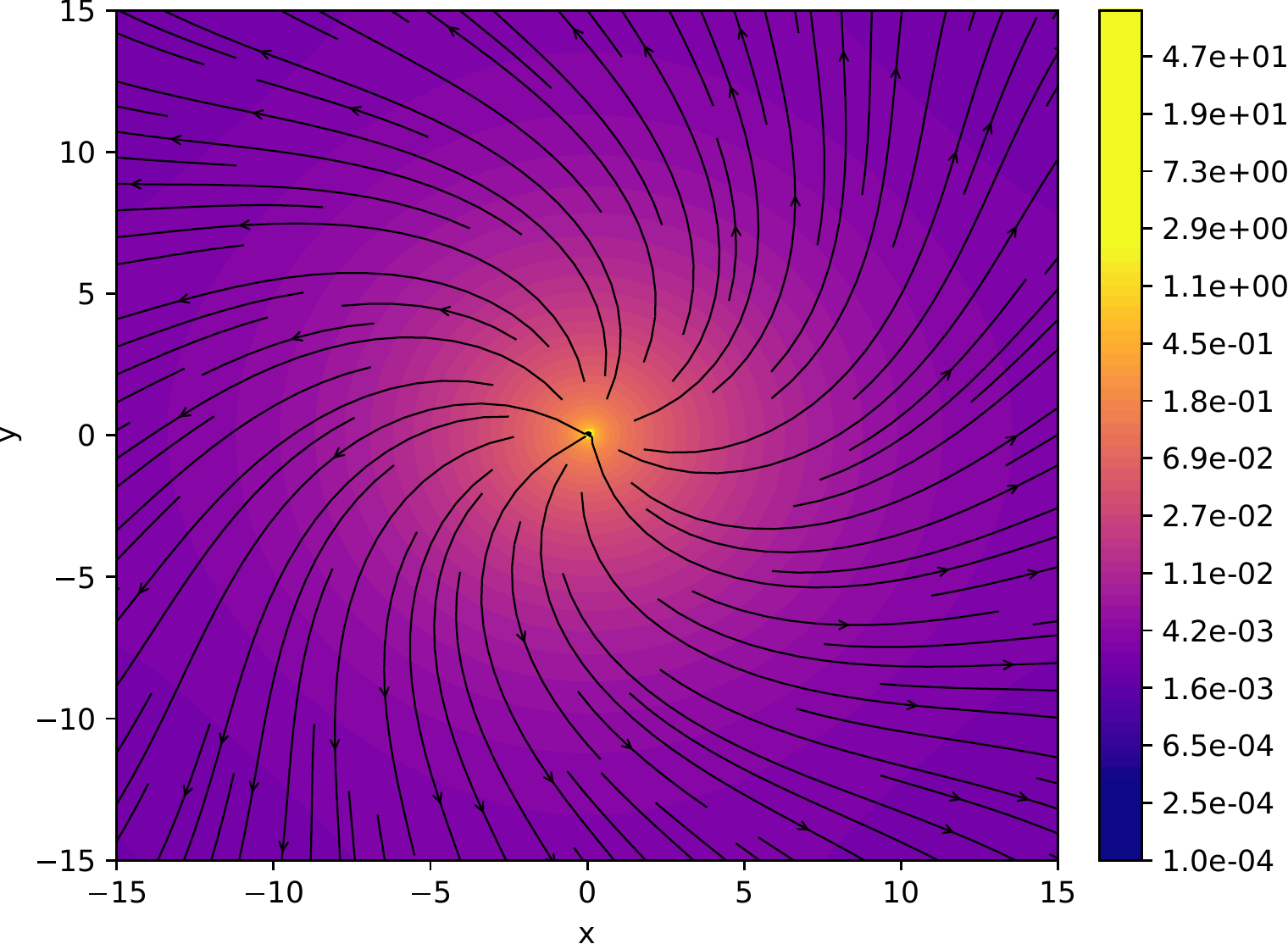}
\caption{
Contour plots of the magnetic field magnitude $|B|$ (colors) in units of $\eta^2$ in the 
$xy-$plane containing the monopole $(z=\zm)$, and the projection of magnetic field lines 
(black curves): $\vec{B} - (\vec{B}\cdot\hat{k})\hat{k}$, for $\gamma=\pi$, $2\zm= 12 \eta^{-1}$.
}
\label{fig:xy twist pi mag fields}
\end{figure}

The magnetic structure of the twisted ($\gamma=\pi$) dumbbell is shown in 
Fig.~\ref{fig:xy twist pi mag fields} and, in marked contrast to the untwisted case, the
magnetic field lines flow away from the dumbbell. The structure agrees with the
$\cos\theta/ r^2$ expression given in \eqref{Brlarger}.
The magnetic field lines in this case tend to pull the monopole and antimonopole apart,
{\it i.e.} they provide a repulsive force between the monopole and antimonopole.

\begin{figure}
\includegraphics[width=0.5\textwidth]{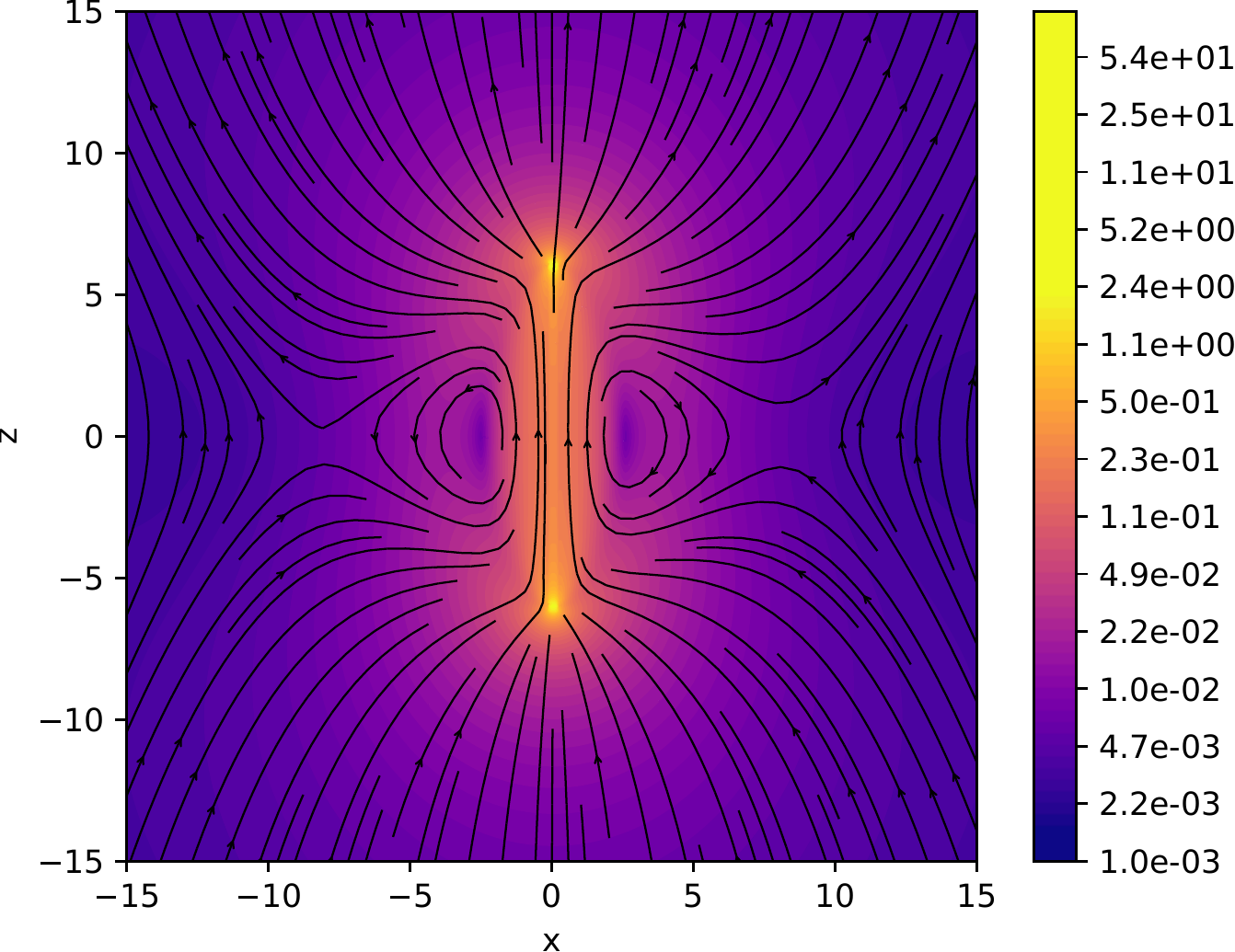}
\caption{
Contours of the magnetic field magnitude $ |B| $ (colors) in units of $\eta^2$, and the projection 
of magnetic field lines in the $xz-$plane after relaxation for $\gamma=\pi$, and
monopole-antimonopole separation $2\zm = 120\delta = 12 \eta^{-1}$.
}
\label{fig:xz twist pi mag fields}
\end{figure}

The field lines in Fig.~\ref{fig:xy twist pi mag fields} show the structure of the magnetic
field projected on to the $xz-$plane: the field lines shown are given by the direction of 
$\vec{B} - (\vec{B}\cdot\hat{y})\hat{y} $ and suppresses the component out of the page
(in the $y-$direction). In Fig.~\ref{fig:xy twist pi mag fields} we show the projected field 
lines in the $z=\zm$ plane. Here we clearly see the twist in the magnetic field lines
first discussed in the context of the sphaleron in Ref.~\cite{Hindmarsh:1994ga,Hindmarsh:1993aw}.

\section{Conclusions and Discussion}
\label{conclusions}

We have developed a numerical technique to study the magnetic structure of electroweak dumbbells in which 
the positions of the monopole and antimonopole are held fixed. We have studied the constrained
solution as a function of the monopole-antimonopole separation and the twist angle. As expected, the
energy grows linearly with separation at large separations, while monopole-antimonopole interactions
become important at small separations. For maximum twist, the dumbbell energy approaches
the electroweak sphaleron energy as the separation goes to zero.

The magnetic field of the electroweak dumbbell at zero twist resembles that of an ordinary bar magnet.
Then the magnetic field strength has the usual dipolar $1/r^3$ fall off at large distances. However the
magnetic field in the case of non-zero twist has an unexpected distribution -- the magnetic field lines
emanating from the monopole, instead of terminating at the antimonopole, are directed towards spatial 
infinity and pull the monopole away from the antimonopole. The magnetic field strength at large
distances has a $\cos\theta/r^2$ fall off. In addition, the magnetic field lines are twisted in the azimuthal 
direction.

For a general electroweak dumbbell formed during electroweak symmetry breaking, the twist angle
will be non-zero and the magnetic field lines emanating from a monopole will terminate on an
antimonopole of some other dumbbell. After the dumbbells have annihilated, the remaining field
lines will perform a random walk in three dimensions and will not close on themselves. This is
likely to have consequences for the correlation length of magnetic fields leftover from the 
electroweak epoch~\cite{Vachaspati:2020blt}. The situation may be similar to that of cosmic strings in
which most of the energy of the cosmic string network is in infinite strings and not in closed loops.
We plan to examine this scenario in more detail in future work.

Another outcome of our work is in the context of Nambu's calculation of the lifetime of
rotating electroweak dumbbells~\cite{Nambu:1977ag}. Our relaxation methods have provided the structure
of the dumbbells which we can feed into an evolution code and study their lifetime as a
function of energy and angular momentum.

\acknowledgements
This work  was supported by the U.S. Department of Energy, Office of High Energy 
Physics, under Award No.~DE-SC0019470.
The authors acknowledge Research Computing at Arizona State University for providing access to high performance computing and storage resources on the Agave and Sol Supercomputer that have contributed to the research results reported within this paper.
\newpage

\bibstyle{aps}
\bibliography{paper}

\end{document}